\newcommand{\dd}{\mathrm{d}}

\documentclass[aps,prl,preprint,groupedaddress]{revtex4-1}
\usepackage{amsmath}
\usepackage{amsfonts}
\usepackage{graphicx}
\usepackage{epstopdf}
\usepackage{caption}
\usepackage{subcaption}
\usepackage{placeins}
\usepackage{hyperref}

\begin{document}

\title{Standard Model vacuum decay with gravity}
\author{Arttu Rajantie}
\affiliation{Department of Physics, Imperial College London, London SW7 2AZ, United Kingdom}
\author{Stephen Stopyra}
\affiliation{Department of Physics, Imperial College London, London SW7 2AZ, United Kingdom}
\date{June 2, 2016}

\begin{abstract}
We present a calculation of the decay rate of the electroweak vacuum, fully including all gravitational effects and a possible non-minimal Higgs-curvature coupling $\xi$, and using the three-loop Standard Model effective potential. Without a non-minimal coupling, we find that the effect of the gravitational backreaction is small and less significant than previous calculations suggested. The gravitational effects are smallest, and almost completely suppressed, near the conformal value $\xi=1/6$ of the non-minimal coupling. Moving $\xi$ away from this value in either direction universally suppresses the decay rate.
\end{abstract}
\pacs{}
\maketitle
\section{Introduction\label{sec:intro}}
Since the discovery of the Higgs boson in 2012 \cite{Aad20121,Chatrchyan201230}, there has been considerable interest in the phenomenon of vacuum decay, motivated by calculations which suggest that the Standard Model effective potential is unstable for the observed value of the Higgs boson and top quark masses  \cite{SHER1989273,Alekhin2012214,Degrassi2012}. The observed masses place the Standard Model firmly in the meta-stability zone, i.e., there exists a second minimum of the potential at much larger field values, to which the electroweak vacuum can decay via the nucleation of bubbles of true vacuum, with the expected life-time of the visible universe longer than its age by many orders of magnitude \cite{Buttazzo2013}. The non-occurrence of such a bubble nucleation event is not inconsistent with this long lifetime, however, the possibility of vacuum decay places constraints on high energy phenomena which might result in the nucleation of true vacuum bubbles. The implications for inflation with a high Hubble rate, for example, have been investigated by many authors \cite{Kobakhidze2013130,Branchina2014,PhysRevLett.112.201801,PhysRevLett.113.211102,PhysRevLett.115.241301}.\\
Lately, there has also been renewed interest in the effect of gravitational back-reaction on vacuum decay in the Standard Model. Calculations show that vacuum decay is dominated by the formation of bubbles with a scale only a single order of magnitude below the Planck scale \cite{Isidori2001387}, where the effect of gravity might be expected to start appearing. The Standard Model effective potential in this regime is negative, and thus the space-time at the center of such a bubble is locally Anti-de-Sitter(AdS), albeit with a sub-Planckian energy density. It was shown by Coleman and de Luccia \cite{PhysRevD.21.3305}, using the thin-wall approximation, that a transition from a zero-energy-density, false vacuum to a negative-energy-density, true vacuum leads to a suppression of the decay rate, due to the warped geometry of the nucleated bubbles. An early paper by Isidori et al.\cite{PhysRevD.77.025034} used a perturbative series in the gravitational coupling to try and estimate the size of this effect. However, it was pointed out recently by Branchina et al.\cite{Branchina:2016bws} that the boundary conditions of the perturbative bounce solution are not satisfied to first order (see discussion section). Additionally, Gregory and Moss et al., in three recent papers \cite{Gregory2014, Burda2015,Burda:2016mou}, investigated vacuum decay in a black-hole background, finding that a black hole can effectively `seed' vacuum decay, in analogy with phase transitions in condensed matter systems.\\
Recent papers have investigated gravitational effects using quartic model potentials; these have the advantage of being easy to compare to the thin-wall results. Branchina et al. \cite{Branchina:2016bws} found that the suppression effect was much less significant, when compared to the thin-wall approximation, for properly treated thick-wall bubbles. Masoumi, Paban, and Weinberg also investigated the thin wall approximation \cite{Masoumi:2016pqb}; they showed that energy arguments for when bubbles may form in the presence of gravitational back-reaction can be extended to thick wall bubbles and also demonstrated that the tunneling rate is not affected by the presence of a Gibbons-Hawking-York boundary term in the action.\\
Another factor, when gravity is included, is the possible presence of a non-minimal coupling between the Higgs field and space-time curvature. It cannot be consistently omitted as, even if zero or near zero at present scales, it will run and become non-zero at higher energies. Therefore it is required for the renormalizability of a scalar field in curved space-time \cite{FREEDMAN197495}. Previously this has been investigated in an inflationary back-ground \cite{PhysRevLett.113.211102}. So far, however, these results have not been extrapolated to the case of a flat space false vacuum. Fortunately, the effect of non-minimal coupling in the present day, with only a small cosmological constant, can be approximated by treating the false vacuum as flat space-time.\\
In this paper we calculate bounce solutions for the Standard Model effective potential, with a non-minimal coupling term $\xi\phi^2R/2$ included in the action. We compute the solutions for the 3-loop Standard Model effective potential by making use of the 3-loop beta functions (with 2 loop pole-matching) available in the literature \cite{Bednyakov2013,Zoller:2014cka}, with which we construct an interpolated potential that can be computed quickly at each step in a numerical integration. We find that the effect of back-reaction on the boundary between instability and meta-stability is small. Furthermore, for a flat space false-vacuum, non-minimal coupling always suppresses Standard-Model vacuum decay relative to the flat space calculation without back-reaction, regardless of the sign of $\xi$. We also find that the conformal value of the coupling, i.e., $\xi = 1/6$ rather than $\xi = 0$, leads to near cancellation of the back-reaction, producing a bounce nearly identical to the flat space bounce. The conformal coupling is found to be a (near) minimum of the decay exponent when $\xi$ is varied, and an exact minimum for the $\lambda\phi^4/4$ potential. We show that if the running of the Higgs coupling did not break the conformal symmetry of the $\lambda\phi^4/4$ potential then this near-cancellation would be exact and the resulting bounces would be identical to flat space. Finally, we present a comparison of our results with the perturbative method of Ref. \cite{PhysRevD.77.025034}; we find that due to the aforementioned failure of the boundary conditions at first order, the perturbative method over-estimates the size of back-reaction effects.

\section{Bounces with Non-minimal Coupling\label{sec:background}}
The decay rate of a metastable vacuum state is given, in the semi-classical approximation, by the Coleman formula \cite{PhysRevD.16.1248}:
\begin{equation}
\Gamma = Ae^{-B},\label{eq:decay_form}
\end{equation}
where $B = S- S_0$ is the difference in Euclidean action between two solutions of the Euclidean field equations of the theory. $S$ is the action of a so-called bounce solution which interpolates between the false and true vacua (though does not in general reach the true vacuum), while $S_0$ is the action of a constant solution sitting at the false vacuum. The pre-factor $A$ is determined, at the semi-classical level, by a functional determinant and is known in the flat space case, see \cite{Isidori2001387}, but has yet to be computed in a curved background. The Euclidean action for a scalar (Higgs) field non-minimally coupled to gravity is:
\begin{equation}
S = \int\dd^{4}x\sqrt{|g|}\left[\frac{1}{2}\nabla^{\mu}\phi\nabla_{\mu}\phi + V(\phi) + \frac{1}{2}\xi\phi^2R - \frac{M_{\rm{P}}^2}{2}R\right],
\end{equation}
where $M_{\rm{P}} = 1/\sqrt{8\pi G_N}$ is the reduced Planck mass. This modifies the original action considered by Coleman and de Luccia \cite{PhysRevD.21.3305} to include a possible non-minimal coupling between the scalar field and gravity. This is required to be present for the theory to be renormalizable \cite{FREEDMAN197495,PhysRevLett.113.211102}. A cosmological constant can also be included, but has here been absorbed into the definition of the potential $V(\phi)$. We neglect boundary terms, which can be dealt with by adding a Gibbons-Hawking-York term to the action \cite{PhysRevD.15.2752}, and do not contribute to the decay rate because the two solutions are identical on the boundary of the Euclidean space-time, if it has a boundary: see Ref. \cite{Masoumi:2016pqb} for a recent discussion of these boundary terms. The resulting Euclidean field equations are:
\begin{align}
R_{\mu\nu} - \frac{1}{2}g_{\mu\nu}R &= \frac{T_{\mu\nu} + \xi\left[-\nabla_{\mu}\nabla_{\nu}\phi^2 + g_{\mu\nu}\nabla_{\lambda}\nabla^{\lambda}\phi^2\right]}{M_{\rm{P}}^2\left(1 - \xi\frac{\phi^2}{M_{\rm{P}}^2}\right)},\label{eq:einstein}\\
T_{\mu\nu}&\equiv \nabla_{\mu}\phi\nabla_{\nu}\phi - g_{\mu\nu}\left[\frac{1}{2}\nabla_{\lambda}\phi\nabla^{\lambda}\phi + V(\phi)\right],\nonumber\\
\nabla_{\mu}\nabla^{\mu}\phi &= V'(\phi) + \xi\phi R.
\end{align}
Note that Einstein's field equations acquire an additional term on the RHS, arising from the $\delta R_{\mu\nu}$ term in the variation of the action which does not form immediately form a total derivative term (as is the case in the variation of the usual Einstein-Hilbert action) due to the position dependent pre-factor $1 - \xi\phi^2/M_{\rm{P}}^2$. For this paper it is assumed that $O(4)$ symmetric solutions dominate the Euclidean action; this has been proven in flat space \cite{ColemanGlaserMartin1978}, but we know of no proof in the curved space case, nor when non-minimal coupling is included. Under this assumption, there is a co-ordinate system in which the metric takes the form:
\begin{equation}
\dd s^2 = \dd\chi^2 + a^2(\chi)\dd\Omega_{3}^2\label{eq:O4metric},
\end{equation}
where $\dd\Omega_{3}^2$ is the unit metric on a 3-sphere. The co-ordinate $\chi$ is the radial distance from the origin, while $a(\chi)$ is the radius of curvature of a 3-sphere at fixed radius $\chi$. Using this co-ordinate system and the assumption of $O(4)$ symmetry, the equations of motion take the form:

\begin{align}
\ddot{\phi} &= -\frac{3\dot{a}}{a}\dot{\phi} + V'(\phi) +\xi\phi R\label{eq:bounce_xi},\\
\dot{a}^2 - 1 &= -\frac{a^2\left[-\frac{\dot{\phi}^2}{2} + V(\phi) - \frac{6\xi\dot{a}\phi\dot{\phi}}{a}\right]}{3M_{\rm{P}}^2\left(1 - \frac{\xi\phi^2}{M_{\rm{P}}^2}\right)},\label{eq:friedmann_first}\\
\ddot{a} &= -\frac{a\left[\dot{\phi}^2 + V(\phi) - 3\xi\left(\dot{\phi}^2 + \phi\ddot{\phi} + \frac{\dot{a}}{a}\phi\dot{\phi}\right)\right]}{3M_{\rm{P}}^2\left(1 - \frac{\xi\phi^2}{M_{\rm{P}}^2}\right)},\label{eq:add}\\
R&= \frac{\dot{\phi}^2(1-6\xi) + 4V(\phi) - 6\xi\phi V'(\phi)}{M_{\rm{P}}^2\left[1 - \frac{\xi(1-6\xi)\phi^2}{M_{\rm{P}}^2}\right]}\label{eq:R},
\end{align}
where dots indicate differentiation with respect to $\chi$. Eq. (\ref{eq:add}) is equivalent to the derivative of Eq. (\ref{eq:friedmann_first}), however, we include it here as it is more reliable for numerics (see Numerical Methods section). The boundary conditions for the bounce are imposed so that the action difference is finite, and depend on the asymptotic behavior of $a(\chi)$:
\begin{itemize}
\item If there exists $\chi_{max} > 0$ such that $a(\chi_{\rm{max}}) = 0$, the Euclidean space-time is compact; we impose $\dot{\phi}(0) = \dot{\phi}(\chi_{\rm{max}}) = 0$ to prevent $\phi$ diverging due to the $1/a$ co-ordinate singularity when $a\rightarrow 0$.
\item If $a(\chi)$  does not cross zero as $\chi\rightarrow \infty$, the Euclidean space-time is non-compact and $\phi(\chi) \rightarrow \phi_{\rm{fv}}$ at infinity (where $\phi_{\rm{fv}}$ is the field in the false vacuum) in order that $S - S_0$ remains finite.
\end{itemize}
The boundary condition $a(0) = 0$, required to solve Eq. (\ref{eq:friedmann_first}), is imposed in both cases. The two scenarios are qualitatively different; the compact space-time of the first gives rise to an effective temperature, and corresponds to a combination of thermal excitation and tunneling in a de-Sitter-like space-time (see \cite{PhysRevD.76.064003} for a discussion). In this paper we focus on the second case, which is the behavior in the Standard Model if the false vacuum is exactly flat space-time.\\
Including the non-minimal coupling, it is possible to simplify the equations by moving to the Einstein frame via a conformal transformation:
\begin{equation}
g_{\mu\nu}\rightarrow \left(1 - \frac{\xi\phi^2}{M_{\rm{P}}^2}\right)g_{\mu\nu}.
\end{equation}
Choosing an analogous co-ordinate system to Eq. (\ref{eq:O4metric}) for the conformally transformed metric gives the bounce equation:
\begin{equation}
\frac{\dd^2\tilde{\phi}}{\dd\tilde{\chi}^2} + 3\frac{\frac{\dd\tilde{a}}{\dd\tilde{\chi}}}{\tilde{a}}\frac{\dd\tilde{\phi}}{\dd\tilde{\chi}} - \frac{\dd}{\dd\tilde{\phi}}\left[\frac{V\boldsymbol{(}\phi(\tilde{\phi})\boldsymbol{)}}{\left(1-\frac{\xi\phi(\tilde{\phi})^2}{M_{\rm{P}}^2}\right)^2}\right] = 0\label{eq:bounceEinstein},
\end{equation}
where
\begin{align}
\dd\tilde{\chi}^2 &= \left(1 - \xi\phi^2/M_{\rm{P}}^2\right)\dd\chi^2\label{eq:chiEinstein},\\
\tilde{a}^2 &= \left(1 - \xi\phi^2/M_{\rm{P}}^2\right)a^2\label{eq:aEinstein},\\
\dd\tilde{\phi} &= \frac{\sqrt{1 - \xi(1-6\xi)\phi^2/M_{\rm{P}}^2}}{\left(1 - \xi\phi^2/M_{\rm{P}}^2\right)}\dd\phi\label{eq:phiEinstein}.
\end{align}
Eq. (\ref{eq:phiEinstein}) can be integrated to obtain $\tilde{\phi}(\phi)$, but the result cannot easily be inverted, analytically, for arbitrary $\xi$. This makes Eq. (\ref{eq:bounceEinstein}) of limited use for numerics, but it can still reveal qualitative features of the solution which are somewhat more opaque in Eq. (\ref{eq:bounce_xi}). Making the variable changes Eqs. (\ref{eq:chiEinstein}) to (\ref{eq:phiEinstein}) transforms Eq. (\ref{eq:bounce_xi}) into (\ref{eq:bounceEinstein}). Furthermore, the critical values of $\phi(0)$ at which the initial (small $\chi$) direction of motion for solutions to Eq. (\ref{eq:bounce_xi}) changes (that is, the false vacuum, barrier peak, and true vacuum of the potential) are determined not by $V(\phi)$ but by:
\begin{equation}
\tilde{V}(\phi) = \frac{V(\phi)}{\left(1 - \xi\phi^2/M_{\rm{P}}^2\right)^2}\label{eq:einstein_potential},
\end{equation}
which is the potential appearing in Eq. (\ref{eq:bounceEinstein}). That this potential determines the initial motion for the solution to Eq. (\ref{eq:bounce_xi}) can be verified by forming the Taylor expansion of the solution for small $\chi$. Note in particular the singularity appearing for $\xi > 0$ at $\phi = M_{\rm{P}}/\sqrt{\xi}$. This is not simply an artifact of the conformal transformation, as it is present in Eq. (\ref{eq:add}). Finite action bounce solutions cannot touch higher field values than this. Since bounces in the Standard Model neglecting gravity are controlled by a scale an order of magnitude below the Planck scale \cite{Isidori2001387}, it is reasonable to conclude that: (a) gravitational back-reaction of the bounce may have an impact, and (b) non-minimal coupling will necessarily distort the peak of the bounce solution. As is clear from Eq. (\ref{eq:einstein_potential}), the effect of non-minimal coupling is negligible if $\phi \ll M_{\rm{P}}/\sqrt{|\xi|}$, but strong above this.

\section{Numerical Methods\label{sec:methods}}
\begin{figure}
\includegraphics[width=0.5\textwidth]{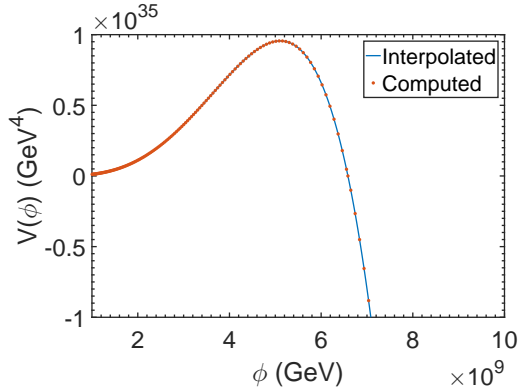}
\caption{\label{fig:sm_potential_interpolation}Interpolated potential for $M_t = 173.34 \text{ GeV}, M_h = 125.15 \text{ GeV}$, compared to ``exact'' values obtained by solving the beta functions numerically.}
\end{figure}

An important numerical challenge for bounce calculations including gravitational effects is that the bounce solutions in a dS-like background do not necessarily touch the false vacuum \cite{Jensen1984176,PhysRevD.76.064003}; this is a property allowed by the compact nature of the Euclidean analogue of dS-like spaces. As a result, large (and in the flat space limit, infinite) contributions to the decay exponent $B = S[\phi_{\rm{bounce}}] - S[\phi_{\rm{fv}}]$ do not cancel explicitly between the bounce and false-vacuum actions (as is the case in the fixed background approximation). Instead a `near cancellation' must occur numerically, which is difficult to study without very high precision calculations. Fortunately, transitions from flat false vacua to AdS true vacua do not suffer from this problem as the Euclidean analogues are non-compact and the bounce solution must touch the false vacuum. It makes sense, therefore, to ignore the small observed positive cosmological constant and consider transitions from a flat false vacuum to AdS-like true vacuum.\\
We use an overshoot/undershoot method to find the bounce solution, as originally proposed by \cite{PhysRevD.21.3305}. Note that we solve Eq. (\ref{eq:add}) rather than Eq. (\ref{eq:friedmann_first}) for the numerics in this paper in order to avoid the square root ambiguity when $\dot{a}$ passes through zero, which happens in a de-Sitter background (for example, in a fixed de-Sitter background with Hubble rate $H$ the solution is $a(\chi) = \sin(H\chi)/H$). We treat Eq. (\ref{eq:friedmann_first}) as a constraint which imposes the second boundary condition ($\dot{a}(0) = +1$) required by Eq. (\ref{eq:add}). This makes no difference to the results for transitions from flat false vacua to AdS true vacua since $\dot{a} > 0$ everywhere for such cases, but it is more generally applicable.\\
To solve the ordinary differential equations (O.D.E.s), Eqs. (\ref{eq:bounce_xi},\ref{eq:add}), we made use of the Odeint library for c++ \cite{odeint}. This library was chosen because it is highly modular in its design and can support a range of different variable types; in particular it naturally supports several types of variable precision numbers. We used the MPFR implementation of the GNU multi-precision library \cite{Fousse:2007:MMB:1236463.1236468,GladmanB} as the variable-precision backend The code uses variable-precision numbers for two reasons: (a) so that it can resolve the vast range of scales (electroweak, barrier scale, $\lambda$ minimization scale, and Planck scale) present in the Standard Model with sufficient precision, and (b) so that it can also be used to study back-reaction in a de-Sitter background. In de-Sitter there is an additional numerical challenge because the boundary conditions do not require $\phi$ to touch the false vacuum; when back-reaction is taken into account, $S_0$ does not cancel analytically in $S - S_0$, as in the fixed back-ground approximation, and since $|S_0| \sim O(M_{\rm{P}}^4/V) \gg |S - S_0|$ generically, rounding errors will wash out the back-reaction if not under stringent control.\\
To approximate the Standard Model potential, we use a piecewise polynomial to interpolate the running of the self-coupling, $\lambda\boldsymbol{(}t(\mu)\boldsymbol{)}$ where $t = \ln(\mu^2/M^2)$. Choosing $\mu = \phi$, the Higgs potential at large scales can be approximated as:
\begin{equation}
V_{\rm{SM}}(\phi) = \frac{\lambda\boldsymbol{(}t(\phi)\boldsymbol{)}\phi^4}{4}.
\end{equation}
We form an approximation of this potential starting from $N+1$ discrete points $\lambda_i = \lambda(t_i), i = 0,1,2,\ldots , N$ obtained by solving the beta functions, which are available in the literature \cite{Buttazzo2013,Zoller:2014cka,Bednyakov2013}, and then fitting $N$ cubic polynomials $\lambda_n(t), n= 1,2,\ldots, N$, each defined only between $t_{n-1}$ and $t_n$:
\begin{align}
\lambda_{n}(t) = &\left[1 - x_n(t)\right]\lambda_{n-1} +   x_n(t)\lambda_{n} + \nonumber\\
&x_n(t)\left[1 -   x_n(t)\right]\left\{a_n\left[1 -   x_n(t)\right] + b_n  x_n(t)\right\},\\
 x_n(t) \equiv& \frac{t - t_{n-1}}{t_{n} - t_{n-1}},\\
a_n =& k_{n-1}(t_{n} - t_{n-1}) - (\lambda_{n} - \lambda_{n-1}),\\
b_n =& -k_{n}(t_{n} - t_{n-1}) + (\lambda_{n} - \lambda_{n-1}).
\end{align}
The constants $k_n$ are the derivatives (with respect to $t = \ln(\phi^2/M^2)$) of the polynomial, at $t_n$. These are chosen to equal the derivatives of $\lambda$, $\dd\lambda/\dd t = \beta_{\lambda}(t)$, resulting in a $C_1$ continuous piece-wise approximation of $\lambda\mathbf{(}t\mathbf{)}$. An alternative choice is to pick $k_n$ such that the piecewise polynomial is $C_2$ continuous; such an approximation is known as a cubic spline \cite{press2007numerical}. However, we found that this led to unwanted oscillation effects in the potential. Figure \ref{fig:sm_potential_interpolation} shows an example interpolated potential compared to the exact values predicted by solving the beta functions. Using this approximation, it is possible to create a model of the Standard-Model potential which is arbitrarily close to the true Standard Model potential simply by taking more initial points to interpolate between. Note that for non-linearly spaced points $t_i$ (e.g., the output of an adaptive O.D.E. solver), selecting the correct polynomial for an arbitrary input $\phi$ may require an interpolative search: since such a search requires on average $O(\rm{log}\boldsymbol{(}\rm{log}(N)\boldsymbol{)})$ steps \cite{Perl1978}, this is not generally a problematic bottleneck.

\section{Results\label{sec:results}}
\begin{figure}
\includegraphics[width=0.5\textwidth]{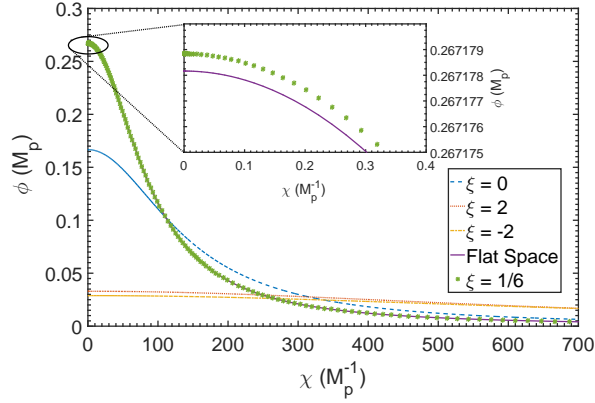}
\caption{\label{fig:various_xi}Bounce solutions as a function of distance $\chi$ from the center of the bounce, computed for various values of $\xi$. The $\xi = 1/6$ solution is extremely close to the flat space solution (though not identical to it). Larger $|\xi|$ tends to flatten and broaden the bounce, as does the inclusion of back-reaction for $\xi = 0$.}
\end{figure}
Figure \ref{fig:various_xi} illustrates a few example bounce solutions for different values of $\xi$, compared with the flat space equivalent. Most notable is $\xi = 1/6$, which is almost (but as the inset plot shows, not quite) identical the flat space case. The boundary value $\phi(0)$ for each bounce is constrained by the overshoot/undershoot method to a range of width $\sim 10^{-15}$, much smaller than the difference between the initial values for $\xi = 0$ and $\xi = 1/6$, while the solutions are computed with an absolute error tolerance of $10^{-20}$ using arbitrary precision variables. This verifies that the small difference between the $\xi = 1/6$ and $\xi = 0$ solutions is not simply a numerical artifact. The effects of positive and negative $\xi$ on the bounce are qualitatively similar.

\begin{figure*}
\includegraphics[width=\textwidth]{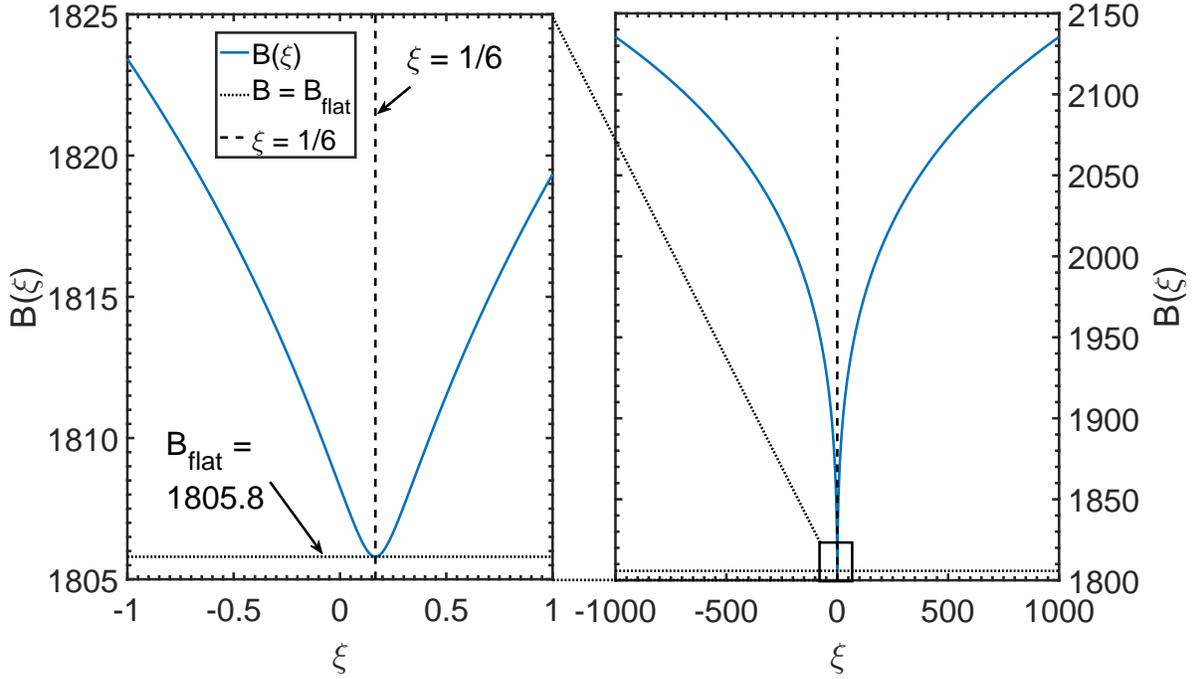}
\caption{\label{fig:xi_range}Decay exponent for different values of $\xi$ at $M_t = 173.34 \text{ GeV}, M_h = 125.15 \text{ GeV}$, compared with the flat space result. Left - variation around the minimum near $\xi = 1/6$. Right - large scale variation of $B(\xi)$}
\end{figure*}
\begin{figure}
\includegraphics[width=0.5\textwidth]{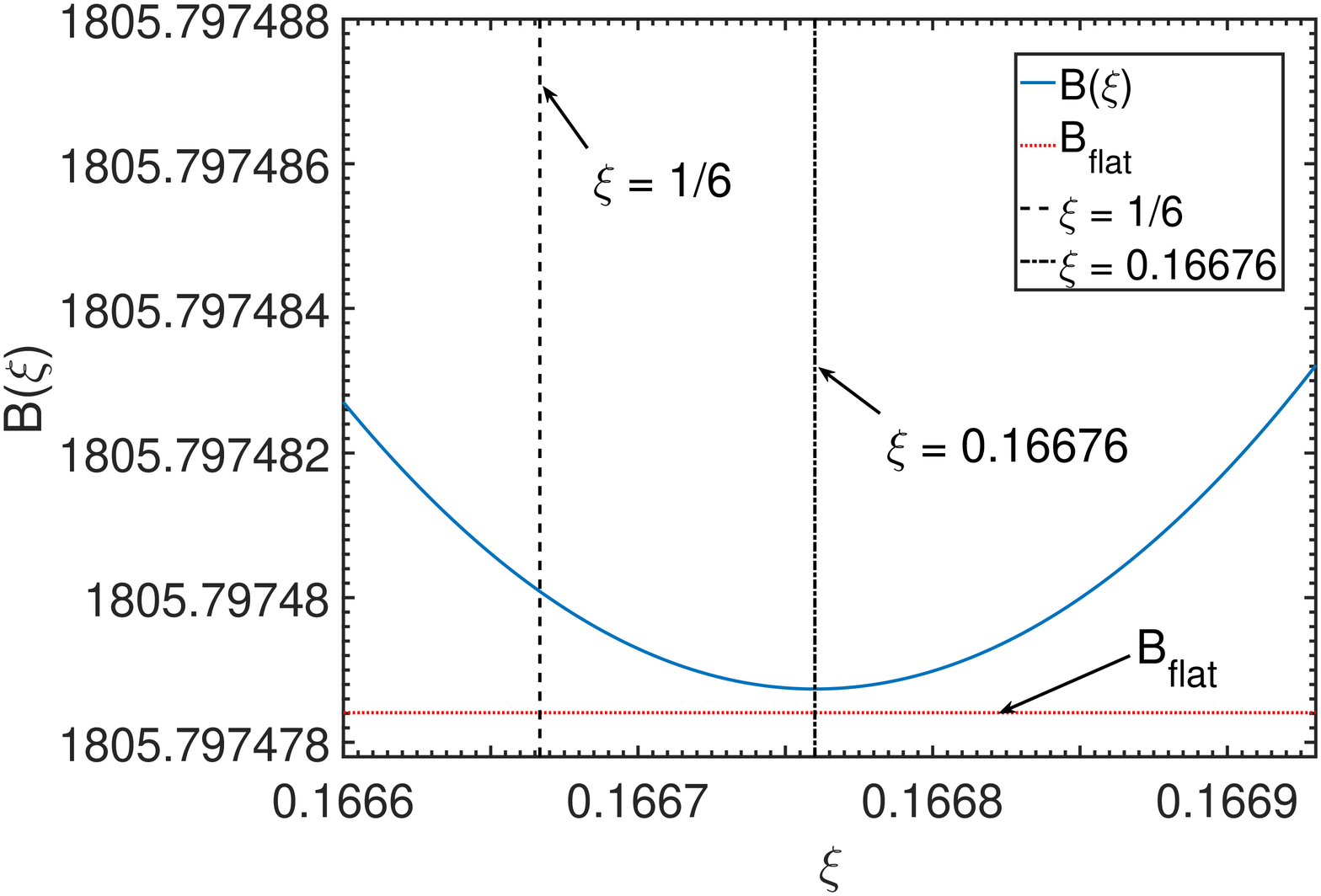}
\caption{\label{fig:xi_range_zoomed} High resolution data around the minimum of $B(\xi)$ for $M_t = 173.34 \text{ GeV}, M_h = 125.15 \text{ GeV}$. This demonstrates that the true minimum is slightly deflected from $\xi = 1/6$ and that the decay, even at the minimum, is suppressed relative to the flat space case.}
\end{figure}
\begin{figure*}
\includegraphics[width=\textwidth]{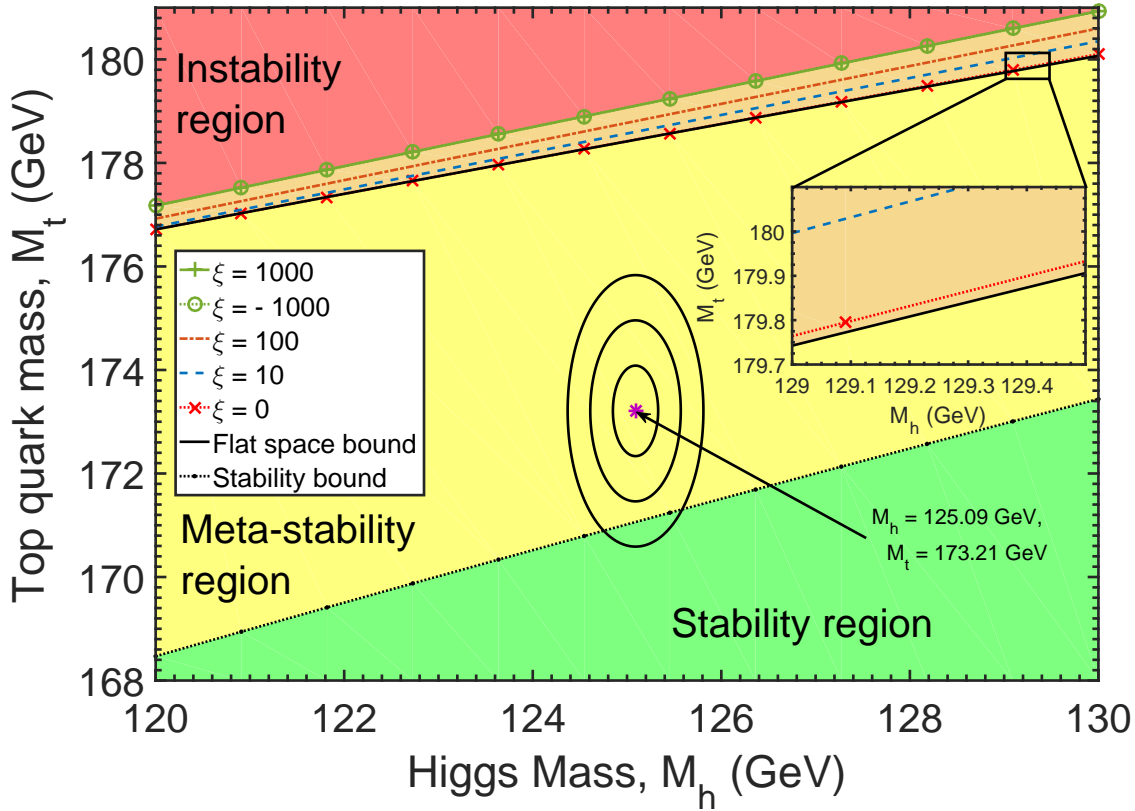}
\caption{\label{fig:stability}Instability/metastability boundaries computed for different values of $\xi$. With $\xi = 0$, the bounds hardly shift at all from the flat space bound (see inset which shows that the nearly coincident flat-space and $\xi=0$ with back-reaction boundaries are in fact separate, albeit very close). With increasing $|\xi|$, the instability region is pushed back; notice that this is true for both positive and negative $\xi$; the boundaries for $\xi = \pm 1000$ are nearly co-incident. The Standard Model parameter uncertainty region around $M_h = 125.09\text{ GeV}, M_t = 173.21 \text{ GeV}$ \cite{Agashe:2014kda} with 1, 2, and 3 sigma bounds is shown for reference.}
\end{figure*}

Figure \ref{fig:xi_range} shows how this affects the decay rate. As with the bounce, the $\xi = 1/6$ case is virtually identical to the flat space case, and represents the approximate minimum decay exponent. For both positive and negative values of $\xi$ away from $\xi = 1/6$, the decay rate is suppressed, increasingly so as $\xi$ is increased. This trend continues to much larger values of $\xi$, until around $\xi\sim 10^{18}$, where the barrier is erased in Eq.  (\ref{eq:einstein_potential}). Figure \ref{fig:xi_range_zoomed} shows the same curve as Fig. \ref{fig:xi_range} but around the minimum, demonstrating the slight deflection from $\xi = 1/6$ and that the curve always lies above the flat space case. To verify that this is not a numerical artifact, this figure is generated using 56642 interpolating polynomials for the running of $\lambda$, with the bounce equations solved at an absolute tolerance of $10^{-20}$ using arbitrary precision variables.\\
We also computed the effect of varying $\xi$ on the boundary between meta-stability and instability. Lacking a complete analysis of the $A$ coefficient in Eq. (\ref{eq:decay_form}), which requires a computation of the functional determinant including graviton loops, we estimate the life-time by assuming $A\sim 1/\bar{R}^4$ where $\bar{R}$ is the full-width-half-maximum (FWHM) of the bounce (this is a good approximation of the quantum corrections in the flat space case \cite{Isidori2001387,PhysRevD.91.013003}). This results in a life-time of \cite{PhysRevD.91.013003}:
\begin{equation}
\frac{\tau}{T_U} = \left(\frac{\bar{R}}{T_U}\right)^4e^{B},\label{eq:lifetime}
\end{equation}
where $T_{U}$ is the age of the visible universe. Although a full analysis of the functional determinant in the flat space case is available \cite{Isidori2001387}, we use the Eq. (\ref{eq:lifetime}) as an estimate in the flat space case in order to separate out the effect of back-reaction alone. As a consequence, our bounds on the meta-stability/instability regions sacrifice accuracy in order to demonstrate the effect of back-reaction. The resulting change in the instability/meta-stability boundary, which we define as the curve where $\tau = T_U$, is plotted in Fig. \ref{fig:stability}.

\section{Discussion\label{sec:disc}}

\subsection{\label{subsec:backreaction}Back-reaction}
Figure \ref{fig:stability} appears to suggest that the effect of gravitational back-reaction on vacuum decay is almost negligible when $\xi = 0$. Even near the instability boundary, the life-time of the vacuum would not shift significantly due to gravitational effects, despite the change in the shape of the bounce. This appears to support the findings of \cite{Branchina:2016bws}, who studied the effect of back-reaction on vacuum instability in a quartic model potential. They found that the thin-wall approximation over-estimated the effect of back-reaction in suppressing vacuum decay compared to the true thick wall bubbles and argued that the inclusion of gravitational back-reaction would not stabilize the potential against decay, even in a strong gravity regime. The bounces we calculated for the actual Standard Model potential are thick-wall bubbles and appear to bear this out; the gravitational back-reaction corrections (for $\xi = 0$) in the Standard Model potential are indeed small. Figure \ref{fig:Ricci} verifies for $\xi = 0$ that the gradient contribution to $R$ is significant in the interior of the bubble; $R$ peaks at the \emph{positive} value $R/M_{\rm{P}}^2 = 2.213\times 10^{-8}$ when $\phi(\chi_{\rm{peak}}) \approx 10^{17} \text{ GeV}$, at which point $V < 0$. This suggests a physical explanation for why the thin-wall approximation over-estimates the suppression of vacuum decay: in the interior of the bubble the thin-wall approximation assumes that the only contribution to $\dot{a}$ is from the true vacuum, i.e.,
\begin{equation}
\dot{a}^2 \approx 1 - \frac{a^2V(\phi_{\rm{tv}})}{3M_p^2}.
\end{equation}
This would correspond to $R < 0$ and so the $R>0$ region of Fig. \ref{fig:Ricci} demonstrates that the approximation has completely broken down. Physically speaking, the back-reaction due to the energy density in the interior of the bubble suppresses the decay rate because it decreases the bubble's volume to surface area ratio compared to flat space (see \cite{PhysRevD.21.3305}); this means an energy-conserving bubble must be larger. As a result, nucleating such a bubble requires over-coming a greater gradient-barrier and so the decay rate is suppressed. However, the back-reaction of the bubble walls, which the thin-wall approximation neglects, counter-acts this effect since for $R > 0$ the opposite is true; the volume-to-surface-area ratio of the bubble increases. The net result is that the interior back-reaction wins, but with a much reduced suppression of the decay rate. This is in addition to the fact that thick-wall bubbles do not touch the true-vacuum, which also reduces $|R|$.\\

\subsection{\label{subsec:non_min_coupling_disc}Effect of non-minimal coupling}
The situation for $\xi \neq 0$ is somewhat different. Most notable is the near cancellation of the back-reaction when $\xi = 1/6$, for which the bounce closely resembles the flat space solution (see Fig. \ref{fig:various_xi}). Figure \ref{fig:xi_range} demonstrates that $\xi = 1/6$ is near the minimal value for the decay exponent; the decay rate is virtually identical to flat space. The reason for this is that $\xi = 1/6$ restores a near conformal symmetry to the bounce equations. Consider the situation where $V(\phi) = \lambda\phi^4/4$, for constant $\lambda < 0$. Such a potential has a conformal symmetry and for $\xi = 1/6$ the equations are, therefore, exactly conformal. The Ricci scalar satisfies Eq. (\ref{eq:R}) and for $\xi = 1/6$ the $\dot{\phi}^2(1-6\xi)$ term vanishes. Additionally, The second and third terms cancel:
\begin{equation}
4V(\phi) - 6\xi \phi V'(\phi) = \lambda \phi^4 - \frac{6}{6}\phi\cdot\lambda\phi^3 = 0.\label{eq:xi_1_6_R_vanishing}
\end{equation}
Thus $R = 0$ holds everywhere for this exactly conformal situation and the back-reaction cancels exactly. It is a simple matter to show that the $\frac{3\dot{a}}{a}$ term in Eq. (\ref{eq:bounce_xi}) reduces to the flat space case as well:
\begin{align*}
R  &= \frac{6(1 - \dot{a}^2)}{a^2} - \frac{6\ddot{a}}{a} = 0 \implies 1 - \dot{a}^2 - \ddot{a}a\\ 
& = 1 - \frac{\dd}{\dd\chi}(\dot{a}a) = 0\\
&\implies \dot{a}a = \chi + C\implies \\
& \frac{1}{2} \frac{\dd}{\dd\chi}(a^2) = \chi + C\implies a(\chi) = \sqrt{\chi^2 + 2C\chi + D},
\end{align*}
where $C$ and $D$ are integration constants. To match the boundary conditions $a(0) = 0, \dot{a}(0) = 1$, it is clear that $a(\chi) = \chi$. Thus, the bounce equations reduce to:
\begin{equation}
\ddot{\phi} + \frac{3}{\chi}\dot{\phi} + |\lambda|\phi^3 = 0.
\end{equation}
This is exactly the flat space bounce equation. Therefore, in the case of $\xi = 1/6$ an analytic expression for the solution can be found in the form of the flat space Lee-Weinberg bounce \cite{Lee1986181}:
\begin{equation}
\phi(\chi) = \sqrt{\frac{2}{|\lambda|}}\frac{2\bar{R}}{(\chi^2 + \bar{R}^2)},\label{eq:LW_bounce}
\end{equation}
where $\bar{R}$ is an arbitrary scale arising due to the exact conformal symmetry. In the case of the Standard Model, the running of $\lambda$ breaks this conformal symmetry. However, the contribution to the Ricci scalar still nearly cancels:
\begin{equation}
4V(\phi) - 6\xi \phi V'(\phi) = -\lambda'(\phi)\frac{\phi^5}{4} = -\frac{\dd\lambda}{\dd \ln(\phi^2/M^2)}\frac{\phi^4}{2};
\end{equation}
it is straightforward to verify numerically that $\dd\lambda/\dd \ln(\phi^2/M^2)$ is small. For example, at $M_h = 125.15 \text{ GeV}, M_t = 173.34 \text{ GeV}$, $|\beta_{\lambda}| = |\dd\lambda/\dd\ln(\phi^2/M^2)| < 1.1\times 10^{-2}$ over the whole range of $\mu$ up to the Planck scale. Also, since the scale of the peak of the bounce is dominated by the scale at which $\beta_{\lambda}$ vanishes, $\beta_{\lambda}$ is by definition tiny in the vicinity of the peak; this is precisely the region when $\phi$ is close to the Planck mass and gravitational effects would matter most. Thus, the bounce for $\xi = 1/6$ in the Standard Model should have $R \approx 0$ and thus be approximately the same as the flat space case. This is precisely what our numerical results show; the bounce is almost identical to the flat space bounce except for a small difference near the peak (see inset Fig. \ref{fig:various_xi}), where the slightly broken conformal symmetry leads to small back-reaction effects.\\

\begin{figure*}
\includegraphics[width=\textwidth]{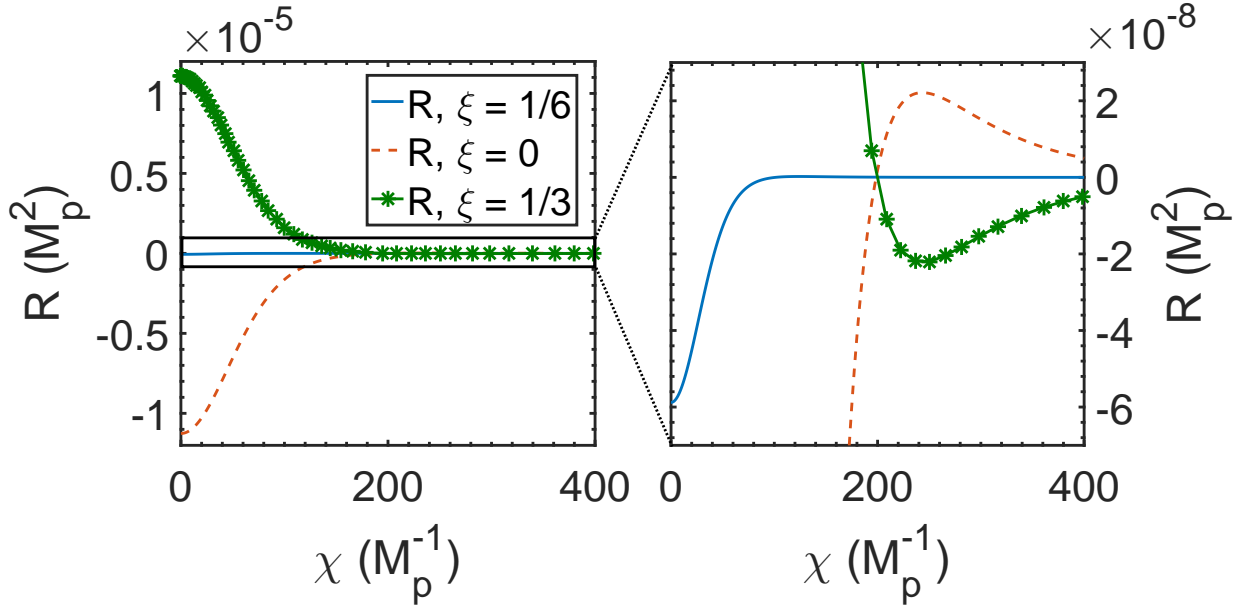}
\caption{\label{fig:Ricci}Ricci scalar with distance from center of bounce, in the $\xi = 0,\xi = 1/6$ and $\xi = 1/3$ cases with $M_t = 173.34\text{ GeV}, M_h = 125.15 \text{ GeV}$. For $\xi = 1/3$, $R$ is positive in the interior of the bounce}
\end{figure*}

To compare the level of back-reaction in each case, we plot the Ricci-scalar for the $\xi = 0,\xi = 1/6,$ and $\xi = 1/3$ cases in Fig. \ref{fig:Ricci}. This demonstrates a significant suppression of the back-reaction, quantified by $R$. For $\xi = 1/6$, the back-reaction is not entirely suppressed, due to the running of $\lambda$ breaking the conformal symmetry.
Away from $\xi = 1/6$, Fig. \ref{fig:xi_range} shows that $B$ increases in both directions; this indicates that the effect of $\xi$ is always to suppress vacuum decay, if the false vacuum is flat. As mentioned in the results section, this effect persists to larger values of $|\xi|$. The degree of suppression is evident in Fig. \ref{fig:stability}, which shows the boundary between stability and instability is pushed back as $|\xi|$ increases (middle region). The effect of pure back-reaction ($\xi = 0$) is also to push back the boundary, although the effect is small.

It is worth noting that since the 3-sigma bounds on the top quark and Higgs masses do not place the Standard Model near the instability boundary, the effect of non-minimal coupling does not qualitatively change the decay behavior of the Higgs field, other than increasing its life-time. In our analysis, the boundary between stability and meta-stability does not change, because whether or not the electro-weak vacuum is false is determined in the Einstein frame by Eq. (\ref{eq:einstein_potential}). With $\xi = 0$ the electro-weak vacuum is stable if $V(\phi)$ is nowhere negative, and meta-stable (or unstable) otherwise. Eq. (\ref{eq:einstein_potential}) does not affect the sign of $V$, nor where it changes sign (the instability scale, $\phi_{\rm{inst}}$), and so the only effect of $\xi$ is to change (in this case decrease) the actual decay rate, if the vacuum is not stable. A completely stable vacuum remains stable when non-minimal coupling is included. Note that this isn't the case in a non-flat back-ground, i.e., $V(\phi_{\rm{fv}})\neq 0$, since there the criterion for stability is that $V(\phi) \geq V(\phi_{\rm{fv}})$ everywhere, rather than $V(\phi) \geq 0$. So, for example, if $\phi_{\rm{fv}} = 0,\xi > 0$ and $V(0) > 0$ is sufficiently large, then $\tilde{V}(0) = V(0)$ and the difference
\begin{equation}
\tilde{V}(\phi) - \tilde{V}(0) = \frac{\Delta V(\phi) + \left[1 - \left(1 - \frac{\xi\phi^2}{M_{\rm{P}}^2}\right)^2\right]V(0)}{\left(1 - \frac{\xi\phi^2}{M_{\rm{P}}^2}\right)^2},
\end{equation}
where $\Delta V(\phi) = V(\phi) - V(\phi_{\rm{fv}})$, can always be made positive for any negative $\Delta V(\phi)$, if $V(0)$ is sufficiently large. In the Jordan frame, this manifests as an extra mass term $\frac{1}{2}\frac{4V(0)}{M_{\rm{P}}^2}\phi^2$ in the potential. For $V(\phi_{\rm{fv}}) = 0$, however, this effect is not present and $\xi > 0$ will not render the false vacuum absolutely stable unless $\xi > M_{\rm{P}}^2/\phi_{\rm{inst}}^2$, i.e., the instability scale $\phi_{\rm{inst}}$ is above the threshold for which a singularity appears in the equations of motion.\\
There is a caveat to the above, however; the stability/meta-stability boundary can in fact shift from the estimate based on the sign of the potential, since in the strong gravity limit it is possible for the gravitational effects of bounces to AdS true vacua to fully quench vacuum decay. This was investigated recently in Ref. \cite{Masoumi:2016pqb}, who found that the quenching effect present in the thin wall approximation persisted for thick-wall bubbles, and when all gravitational effects were taken into account. We do not investigate this effect here, but it can in principle lead to shifts in the stability/meta-stability boundary.

\subsection{\label{subsec:exact_min}Exact minimum under $\xi$ variation}
The form of Fig. \ref{fig:xi_range} appears to suggest a minimum near to $\xi = 1/6$. Assuming that $B(\xi)$ and $\phi_{\xi}(\chi)$ (the bounce solution with non-minimal coupling $\xi$) vary smoothly with $\xi$, we can construct an analytic expression for $B'(\xi)$. Recall that $B = S - S_{\rm{fv}}$ where $S_{\rm{fv}}$ is the false vacuum action. Assuming that the false vacuum lies at $\phi = 0$ for all $\xi$, then $S_{\rm{fv}}$ is independent of $\xi$ because $R = 0$ for that solution (this would not be the case for a transition from a de-Sitter false vacuum, however). Consequently, we can write
\begin{align}
\frac{\dd B}{\dd \xi} =& \left.\frac{\partial S_{\xi}[\phi_{\xi},g_{\xi,\mu\nu}]}{\partial\xi}\right|_{\phi,g_{\mu\nu}} + \left.\frac{\delta S_{\xi}[\phi_{\xi},g_{\xi,\mu\nu}]}{\delta \phi}\right|_{\xi,g_{\mu\nu}}\frac{\partial \phi_{\xi}}{\partial \xi} + \nonumber\\ &\left.\frac{\delta S_{\xi}[\phi_{\xi},g_{\xi,\mu\nu}]}{\delta g_{\xi}^{\mu\nu}}\right|_{\xi,\phi}\frac{\partial g_{\xi}^{\mu\nu}}{\partial \xi}
\end{align}
where $\phi_{\xi}$ and $g_{\xi,\mu\nu}$ are the scalar field and metric bounce solutions respectively, when the action is $S_{\xi}$ defined by
\begin{align}
&S_{\xi}[\phi,g_{\mu\nu}] \equiv\nonumber\\
&\int\dd^4x\sqrt{g}\left[\frac{1}{2}\nabla_{\mu}\phi\nabla^{\mu}\phi + V(\phi) + \frac{1}{2}\xi\phi^2 R- \frac{M_{\rm{P}}^2}{2}R\right].
\end{align}
Defining a new functional
\begin{equation}
\left.\frac{\partial S_{\xi}}{\partial\xi}\right|_{\phi,g_{\mu\nu}} = \Delta S[\phi,g_{\mu\nu}] \equiv \int\dd^4x\frac{1}{2}\phi^2 R,
\end{equation}
and using the fact that the first functional derivatives of $S_{\xi}$ at $\phi_{\xi},g_{\xi,\mu\nu}$ vanish (as these are stationary points of $S_{\xi}$), we find:
\begin{equation}
\frac{\dd B}{\dd \xi} = \pi^2\int_{0}^{\infty}\dd\chi a_{\xi}^3(\chi)\phi_{\xi}^2(\chi)R_{\xi}(\chi).\label{eq:dBxi}
\end{equation}
All the quantities in the integrand can be evaluated from the bounce solution, once it is found. We confirmed that this prediction for the derivative of $B(\xi)$ agrees with numerical differentiation of the data in Fig. \ref{fig:xi_range}. Eq. (\ref{eq:dBxi}), however, yields useful analytic insight into the shape of Fig. \ref{fig:xi_range}. In particular, it reveals that $\xi = 1/6$ would be the exact minimum of the potential for a constant $\lambda$ quartic potential, since Eq. (\ref{eq:xi_1_6_R_vanishing}) shows that the Ricci scalar vanishes everywhere for $\xi = 1/6$, hence $B'(\xi) = 0$. In the Standard Model, the conformal symmetry of the large scale Higgs potential is broken by the running of $\lambda$, which manifests in the non-vanishing of $R$ (see Fig. \ref{fig:Ricci}). Consequently, the minimum shifts to a value slightly different from $\xi = 1/6$. The near vanishing of $R$ for $\xi = 1/6$, however, explains the observation that Fig. \ref{fig:xi_range} has a minimum very close to $\xi = 1/6$. The exact minimum can be found by a root finding algorithm using Eq. (\ref{eq:dBxi}). For example, if $M_h = 125.15 \text{ GeV}$ and $M_t = 173.34 \text{ GeV}$ the minimum lies at $\xi_{\text{min}} = 0.16676$, a shift of  $\xi_{\text{min}} - 1/6 = 9.3354\times 10^{-5}$.

\subsection{\label{subsec:comparison}Comparison to previous results}
\begin{figure*}
\includegraphics[width=\textwidth]{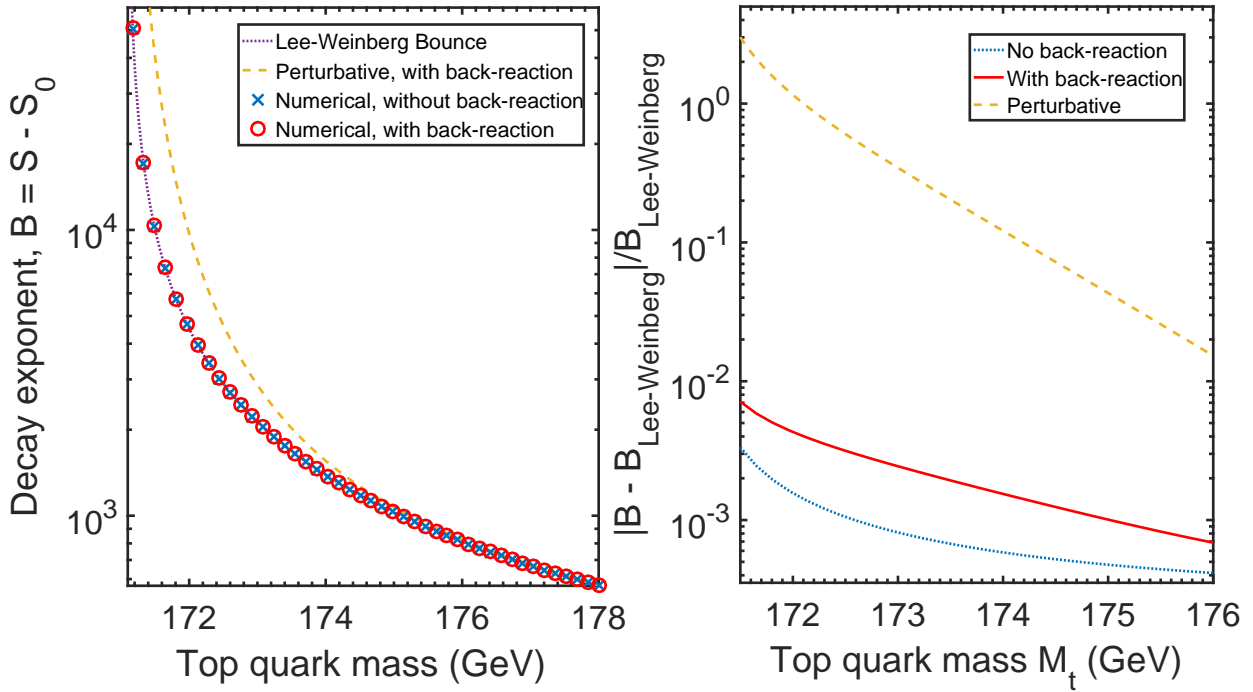}
\caption{\label{fig:comparison_isidori}Left: comparison of the perturbative results of Ref. \cite{PhysRevD.77.025034} to our numerical calculations in the full Standard Model potential, for $M_{h} = 125.15 \text{ GeV}$. The numerical calculations show only a small shift from the flat space, constant-$\lambda$, Lee-Weinberg result, while the perturbative formula predicts a much stronger suppression. Right: fractional difference between the decay exponents on the left and the Lee-Weinberg prediction. Back-reaction is found to increase the action, but the fractional effect is small, and much less significant than the perturbative series predicts. Note - the numerical calculation of the flat space action does not exactly match the Lee-Weinberg prediction as it is computed in the 3-loop quantum corrected potential, rather than the constant $\lambda$ potential.}
\end{figure*}

Ref. \cite{PhysRevD.77.025034} previously computed an analytic correction to the bounce action via a perturbation expansion in the gravitational coupling, $\kappa = \frac{1}{M_{\rm{P}}^2}$, or more precisely, the dimensionless quantity $1/(\bar{R}^2M_{\rm{P}}^2)$ where $\bar{R}$ is the bounce length-scale - see Eq. (\ref{eq:LW_bounce}):
\begin{align}
\phi(\chi) &= \phi_{0}(\chi) + \kappa\phi_1(\chi) + O(\kappa^2),\label{eq:phi_pert}\\
a(\chi) &= a_0(\chi) + \kappa a_1(\chi) + O(\kappa^2).
\end{align}
where $a_0(\chi) = \chi$, and $\phi_0$ is the flat space bounce in the constant $\lambda$ potential (which is more convenient to quantize around). Note for comparison that we define $M_{\rm{P}}^2 = 1/8\pi G_N$, whereas Ref. \cite{PhysRevD.77.025034} uses $M_{\rm{P}}^2 = 1/G_N$. However, it was recently pointed out (\cite{Branchina:2016bws}) that this perturbation expansion fails to satisfy the boundary conditions at first order in $\kappa$. This can be demonstrated by considering the equations for the first order perturbations. Substituting the expansions into the bounce equations we obtain, comparing the first order in $\kappa$,
\begin{align}
\ddot{\phi_1}&= - \frac{3}{\chi}\left(-\frac{a_1(\chi)}{\chi}\dot{\phi_0} + \dot{\phi_1} + \dot{a_1}\dot{\phi_0}\right) + V''(\phi_0)\phi_1,\\
\dot{a_1} &= \frac{\chi^2}{3}\frac{8\bar{R}^2}{|\lambda|}\frac{1}{(\chi^2 + \bar{R}^2)^3},\\
\phi_0(\chi) &= \sqrt{\frac{2}{|\lambda|}}\frac{2\bar{R}}{\bar{R}^2 + \chi^2},  \dot{\phi_0} = -\sqrt{\frac{2}{|\lambda|}}\frac{4\bar{R}\chi}{(\bar{R}^2 + \chi^2)^2}\nonumber.
\end{align}
The equation for $a_1$ can be integrated immediately. Setting $x = \chi/\bar{R}$ and $y = \bar{R}^3|\lambda|^{3/2}\phi_1$ we find the equation for the first order perturbation of the field is
\begin{align}
&y''(x) + \frac{3}{x}y'(x) + \frac{24}{(1 + x^2)^2}y + \tilde{f}(x) = 0,\\
&\tilde{f}(x) \equiv 4\sqrt{2}\left(\frac{(x^2 - 1)}{(1 + x^2)^4} + \frac{\text{arctan}(x)}{x(1 + x^2)^2} - \frac{8 x^2}{(1 + x^2)^5}\right)\nonumber.
\end{align}
The general solution to this equation is
\begin{align}
y(x) =& \frac{C_1(x^2 - 1)}{(1 + x^2)^2} \nonumber\\
&+ \frac{C_2[1 -17 x^2 - x^2(1 - x^2)\left(x^2 + 12\ln x\right)]}{2x^2(1 + x^2)^2}\nonumber\\& + \frac{4\sqrt{2}}{45x^2(1 + x^2)^3}\left[1 + x^2(7 - 9x^2)\vphantom{\int}\right.\nonumber\\
&\left. - 15x^3(1 + x^2)\text{arctan}(x) + 6x^2(1 - x^4)\ln\left(\frac{1 + x^2}{x^2}\right)\right].\label{eq:GS}
\end{align}
In the limit as $x\rightarrow\infty$ one finds $y(x) \rightarrow C_2/2$. When $x\rightarrow 0$ however, we obtain the asymptotic form
\begin{align}
y(x) \rightarrow & \left(\frac{C_2}{2} + \frac{4\sqrt{2}}{45}\right)\left(\frac{1}{x^2} - 12\ln x\right) + \nonumber\\
&\left(-C_1 - \frac{19}{2}C_2 + \frac{16\sqrt{2}}{45}\right) + O(x^2\ln x).
\end{align}
Note that $\phi_0$ satisfies $\dot{\phi}_0(0) = 0, \phi_0(\chi\rightarrow\infty) \rightarrow 0$, and $\phi(\chi)$ is required to satisfy the same boundary conditions: this implies that $\phi_1(\chi)$ (and therefore $y(x)$, which is related to $\phi_1$ by a constant factor) must satisfy the same conditions. Thus $y'(0) = 0$ and $y(x\rightarrow\infty)\rightarrow 0$ are required. It is clear from the two limits of Eq. (\ref{eq:GS}) that these boundary conditions cannot be simultaneously satisfied. If $y(x)$ is regular at $x = 0$ then it tends to $-4\sqrt{2}/45$ as $x\rightarrow\infty$ and if it tends to zero (the false vacuum) at $x \rightarrow \infty$ then it will fail to be regular at $x = 0$. This indicates that the perturbation expansion, Eq. (\ref{eq:phi_pert}), always breaks down for some values of $\chi$ and hence the solution cannot be trusted. The perturbative expansion predicts the decay exponent to be
\begin{equation}
B = \frac{8\pi^2}{3|\lambda(\mu)|} + \frac{96\pi^2}{135M_p^2\bar{R}^2|\lambda(\mu)|^2},\label{eq:isidori_B}
\end{equation}
where $\mu$ is the scale at which $B$ is minimized. We neglect the quantum corrections from the functional determinant so as to better compare this formula with our own results. The authors of Ref. \cite{PhysRevD.77.025034} argue that $B$ is independent of $\phi_1$, depending only on $a_1$, to first order in $\kappa$ because the relevant term in the action vanishes when the equations of motion for $\phi_0$ are satisfied. Nevertheless, our numerical results are in conflict with Eq. (\ref{eq:isidori_B}), which over-estimates the suppressing effect of back-reaction on the decay exponent (in practical terms, this does not make a significant difference to the stability bound as the effect is still small). Our results give a decay exponent which is much closer to the flat space result (see Fig. \ref{fig:comparison_isidori}).

\FloatBarrier
\section{Conclusion\label{sec:conc}}
We have computed the full gravitational effects on bounces in the Standard Model. For the region of interest, the effect on the stability bounds of pure ($\xi = 0$) back-reaction is almost negligible. Our results disagree with those of Ref. \cite{PhysRevD.77.025034}, which we find over-estimates the size of gravitational effects. Our calculation supports the recent prediction by the authors of Ref. \cite{Branchina:2016bws} that the gravitational back-reaction should not significantly suppress vacuum decay.\\
Furthermore, we have computed, for what we believe is the first time, the effect on the decay rate of including a non-minimal coupling term in the action. We found that gravitational effects suppress vacuum decay universally for all values of the non-minimal coupling and push back the boundary between instability and stability. The effect, however, would not stabilize the potential completely because it does not change the location of the boundary between the stable/meta-stable regions of $(M_h,M_t)$ phase space. Our calculations singled out the conformal value of $\xi = 1/6$ as being particularly special. For this value of the non-minimal coupling the effect of back-reaction was found to nearly cancel, with the cancellation failing to be exact due to the running of the Higgs self coupling, $\lambda(\mu)$, which breaks the conformal symmetry. $\xi = 1/6$ is also a near minimum of the decay exponent as a function of $\xi$. We showed that in a exactly conformal potential with $\xi = 1/6$: (a) the back-reaction completely cancels, (b) the bounce solution is identical to the flat space bounce, and (c) the decay exponent is minimal under $\xi$ variations. Each of these properties is found to nearly hold in the full Standard Model potential, as it is `nearly' conformal at large field values.\\
Our results show that minimal-coupling gravitational effects in computing Standard Model vacuum decay rates are small, and provide a first analysis of the impact of non-minimal coupling. There is still much to be understood about the effect of gravity and non-minimal coupling on vacuum decay, however. For example, our results do not take account of the running of $\xi$, which should be present in a complete description of its effect on vacuum decay. Nor do we take into account the effect of graviton loops on the running of the Standard Model couplings. A proper analysis of these issues will require a study of the quantum corrections to the gravitational bounce in the form of the functional determinant pre-factor, $A$, in Eq. (\ref{eq:decay_form}). Additionally, as mentioned in the discussion, there may be a shift in the stability/metastability boundary associated with the quenching effect of AdS true vacua\cite{Masoumi:2016pqb}. The nature of this effect when non-minmal coupling is included is a natural direction for further investigation. Finally, the role of non-minimal coupling in non-flat back-grounds, for example during inflation, is also an important avenue for further study because it is possible for non-minimal coupling to stabilize the Standard Model potential if $V(\phi_{\rm{fv}}) > 0$ \cite{PhysRevLett.113.211102}. Such an analysis could therefore provide useful insight into the implications of vacuum meta-stability for early universe physics.
\subsection{Acknowledgments}
We thank Erick Weinberg and Tommi Markkanen for useful discussions. AR was supported by STFC grant ST/L00044X/1 , and SS by the Imperial College PhD Scholarship.

\bibliography{vacuum_decay_gravitational_back_reaction_references}

\end{document}